\documentclass[a4paper]{article}

\usepackage{INTERSPEECH2021}
\usepackage{graphicx}
\usepackage{mathrsfs}
\usepackage{multirow}
\usepackage{tabularx}
\usepackage{lipsum}
\usepackage{cite}

\usepackage{xcolor}

\usepackage{amsmath,multirow,array,enumitem,adjustbox}
\usepackage{subcaption,booktabs}
\usepackage{cleveref}

\title{Improving Non-native Word-level Pronunciation Scoring with Phone-level Mixup Data Augmentation and Multi-source Information}
\name{Kaiqi Fu, Shaojun Gao, Kai Wang, Wei Li, Xiaohai Tian, Zejun Ma}

\address{
  Bytedance AI-Lab, China}
\email{\{fukaiqi.z, gaoshaojun, wangkai.kw, liwei.speech, xiaohai.tian, mazejun\}@bytedance.com}

\begin{document}

\maketitle

\begin{abstract}
Deep learning-based pronunciation scoring models highly rely on the availability of the annotated non-native data, which is costly and has scalability issues.
To deal with the data scarcity problem, data augmentation is commonly used for model pretraining. In this paper, we propose a phone-level mixup, a simple yet effective data augmentation method, to improve the performance of word-level pronunciation scoring.
Specifically, given a phoneme sequence from lexicon, the artificial augmented word sample can be generated by randomly sampling from the corresponding phone-level features in training data, while the word score is the average of their GOP scores.
Benefit from the arbitrary phone-level combination, the mixup is able to generate any word with various pronunciation scores. Moreover, we utilize multi-source information (e.g., MFCC and deep features) to further improve the scoring system performance.
The experiments conducted on the Speechocean762 show that the proposed system outperforms the baseline by adding the mixup data for pretraining, with Pearson correlation coefficients (PCC) increasing from 0.567 to 0.61. The results also indicate that proposed method achieves similar performance by using 1/10 unlabeled data of baseline. 
In addition, the experimental results also demonstrate the efficiency of our proposed multi-source approach.
%\blfootnote{First two authors contributed equally.}

\end{abstract}

\section{Introduction}

It is well-known that the L2 learning process is heavily affected by a well-established habitual perception of phonemes and articulatory motions in the learners' primary language (L1) \cite{ellis1994study}, which often cause mistakes and imprecise articulation in speech productions by the L2 learners, e.g., a negative language transfer \cite{ellis1994study,meng2009developing}. 
As a feasible tool, computer assisted pronunciation training (CAPT) is often employed to automatically assess L2 learners’ pronunciation quality at different levels, e.g., phone-level \cite{franco1997automatic,witt2000phone,zheng2007generalized,hu2015improved,shi2020context,li2017improving,leung2019cnn,yan2020end,feng2020sed,wei2009new}, word-level \cite{qian2012use,kyriakopoulos2020automatic,lee2013pronunciation,lee2016language,korzekwa2021weakly} and sentence-level \cite{cincarek2009automatic,yu2015using,chen2018end,lin2020automatic,lin2021deep}.

Over the last few decades, many methods have been proposed to evaluate non-native leaners' pronunciation quality at the aforementioned levels. Most of these methods can be divided into two major groups, namely one-step or two-step approaches. Methods in the former group directly utilize confidence scores \cite{franco1997automatic,witt2000phone,zheng2007generalized,hu2015improved,shi2020context} derived from automatic speech recognition (ASR) systems to assess phone-level and sentence-level pronunciation quality. For example, the HMM-based phone log-likelihood \cite{franco1997automatic} or log-posterior \cite{witt2000phone,zheng2007generalized,hu2015improved,shi2020context}. Among these methods, Goodness of Pronunciation (GOP)~\cite{witt2000phone} is widely used in pronunciation scoring task. Given an acoustic segment of phoneme with multiple frames, the corresponding GOP is defined as the normalized frame-level posterior probability over the segment.
% As a well-known confidence method, goodness of pronunciation (GOP) score \cite{hu2015improved} is defined to be the duration normalized of posterior probability that the speaker uttered phone given the corresponding acoustic segment.
%Goodness of Pronunciation (GOP) is one of the most widely adopted confidence scores in pronunciation scoring task, w
%As a well-known confidence method, goodness of pronunciation (GOP) score \cite{hu2015improved} is defined to be the duration normalized of posterior probability that the speaker uttered phone given the corresponding acoustic segment.
%Subsequently, the average of the individual confidence scores over the $N$ phones in target sentence is calculated as sentence-level score \cite{franco1997automatic}.

%*************************
%Goodness of Pronun- ciation (GOP) [5] is one of the the most widely adopted feature in speech evaluation task

% For example, the HMM-based phone log-likelihood \cite{franco1997automatic} or log-posterior (e.g., Goodness of pronunciation (GOP) score) was adopted in \cite{franco1997automatic,witt2000phone,hu2015improved} as a confidence score to measure the quality of pronunciation for the target phone. 
% Subsequently, the average of the individual confidence scores over the $N$ phones in target sentence is calculated as sentence-level score \cite{franco1997automatic}.

In addition, the output phone sequence from ASR, e.g. end-to-end models \cite{yan2020end,leung2019cnn,feng2020sed} or extended recognition network \cite{qian2012use,kyriakopoulos2020automatic}, can be aligned with the given prompts, and a phone or word is marked as mispronounced if there is some mismatch between these two sequences. Alternatively, the two-step approaches treat pronunciation scoring or mispronunciation detection as regression or classification task. Specifically, phone, word and sentence boundaries are first generated by forced-alignment, and then either frame-level or segmental-level pronunciation features within each boundary are fed into task-dependent classifiers or regressors (e.g., \cite{hu2015improved,wei2009new,li2017improving,lee2013pronunciation,lee2016language,korzekwa2021weakly,cincarek2009automatic,lin2021deep,lin2020automatic,yu2015using,chen2018end}). Finally, the posterior probabilities or predicted values obtained from those models are often used as pronunciation scores. Taking advantage of using extra supervised training data, the two-step approach armed with deep learning techniques was reported to achieve better scoring performance than its counterparts \cite{lee2016language}.

Although the aforementioned two-step approaches have achieved satisfactory pronunciation scoring or mispronunciation detection results, the performances heavily rely on the size of labeled scoring samples, that is not always practical in many real-world applications. Meanwhile, the non-native data collection and labeling process is costly and has scalability issues \cite{lee2016language}. Take the recently released public free dataset Speechocean762 \cite{zhang2021speechocean762} for example, only 5,000 sentences, 31,816 words and 95,078 phones have been assigned with human scores. Another public dataset L2-ARCTIC \cite{zhao2018l2} has collected 11,026 utterances including 97,000 words and 349,000 phone segments. The comparison in \cite{nancy} show that the largest non-native corpus contains 90,841 utterances, but it is not publicly available. To tackle the limited non-native data problem, data augmentation techniques have been investigated in some previous work \cite{lee2016language,korzekwa2020detection,lin2021deep}. In \cite{korzekwa2020detection}, Text-To-Speech (TTS) was used to synthesize ``incorrect stress" samples on top of modified lexical stress. In \cite{lee2016language}, the author created negative training samples by modifying the canonical text as the modified text forms a mismatched pair with its original word-level speech. More recently, authors in \cite{lin2021deep} collected a large number of unlabeled non-native data for pretraining sentence-level scorer with GOP scores. It avoids time-consuming labeling, but non-native data collection is still a challenge task.

\begin{figure*}[t]
\centering
\includegraphics[width=0.9\textwidth]{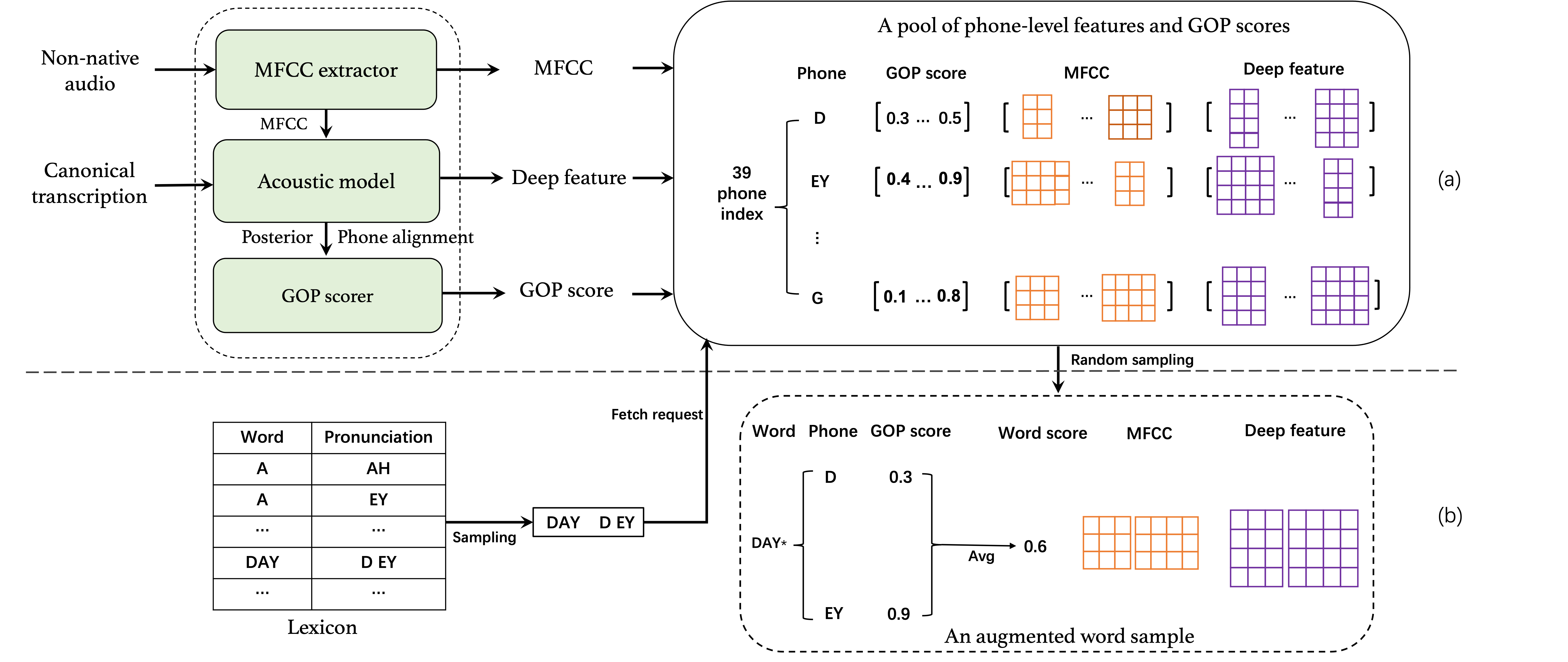}
\caption{The mixup data generation module.}
\label{fig:mix}
\end{figure*}

%Inspired by the mixup in image classification \cite{mixup}, 
%Inspired by the recent success of mixup on image classifications \cite{mixup} and NLP tasks \cite{mixtext}, 
Inspired by the recent success of mixup on image classifications \cite{mixup}, this paper proposes a phone-level mixup to deal with limited word-level training data. Different to image, word pronunciation can be explicitly defined as a sequence of phones. Hence, we are able to build a phone-level dataset from the limited word-level training samples. %Subsequently, a large amount of word-level augmented training samples can be created by sampling some data points in phone-level training set and averaging their corresponding scores as word-level scores. 
For each phone, a pool of phone-level features and corresponding GOP scores is first constructed from the training data. Then the artificial word-level augmented features are concatenated by random sampling from the corresponding phone-level data, where the word score is the average of the phone GOP scores.
Leveraging the arbitrary combination of phone-level information, we can generate any word with various pronunciation scores to improve model's generalization ability. As reported in \cite{li2017improving}, deep features extracted from non-native acoustic model are not discriminate enough due to the inconsistency of non-native phone label. Therefore, we used both MFCC and deep features to further improve the scoring system. 

\section{Pronunciation Scoring Framework with Mixup}
\label{sec:method}
%Our proposed scoring framework contains two steps: (i) the scoring model pretraining process using augmented data, which combines the real speech and artificial words generated by mixup, and (ii) the scoring model fine-tune process using the human labeled data.
Our proposed scoring framework contains two modules: the mixup data generation module, which is able to create arbitrary words with various pronunciation scores, and the pronunciation scorer module, which is used to map a sequence of frame-level acoustic features into word-level pronunciation score.

\subsection{Mixup Data Generation}
\label{ssec:mixup}
Mixup was first proposed in \cite{mixup}, where interpolation between original feature-target pairs were used to generate augmented data for image classification, where the linear interpolation is applied on the pixel-based feature space. 
Different to the image data used in classification task, which has the same size and inherent ordering, for the word-level pronunciation scoring, the speech signal generally consists of different phoneme sequences and various duration, which is hard to perform interpolation on word-level features directly.
Alternatively, we apply mixup on a lower-level feature space, phone-level features. Given a phone sequence from lexicon, it is easy to generated the augmented word-level features by concatenating the corresponding phone-level data, while the word score is the average of the phone scores.

To perform mixup, we need to collect a pool of phone-level features and their corresponding GOP scores. Specifically, an acoustic model, trained with paired speech signal and transcription, is utilized for deep features and GOPs score extraction. As shown in Figure~\ref{fig:mix} (a), given a collection of non-native audio with canonical transcription, we first extract MFCC features from the speech signal. An acoustic model is utilized to force-align all non-native audio (pairs of MFCC features and canonical transcription) to get the start and end time boundary for each phoneme. The MFCCs of each phoneme are then fed into the acoustic model to extract corresponding deep features and frame-level posterior probability from the bottleneck layer~\cite{lin2021deep} and output layer, respectively. Substantially, the GOP score of a specific phoneme can be interpreted as the normalized posterior probability over the phoneme duration obtained by the force-alignment.
Finally, a pool of phone-level features and corresponding GOP scores is collected for each phoneme.
%GOP of target phone $p$ is defined to be the duration normalized of posterior probability that the speaker uttered phone $p$ given the corresponding acoustic segment.

The constructed pools are then used for mixup as shown in Figure~\ref{fig:mix} (b). Firstly, a word and its phoneme sequence is sampled from lexicon based on the word frequency in training set. For each phoneme in the sequence, we randomly sample one quadruplet from the pool of the corresponding phoneme. As the example shown in Figure \ref{fig:mix}, given a word $\mathrm{DAY}$ sampled from the lexicon, we first obtain its phoneme sequence of /D EY/. Then for each phoneme /D/ and /EY/, the GOP scores and corresponding features are randomly sampled from the pool of /D/ and /EY/, respectively.
Finally, the features of generated word is formed by concatenating the features of /D/ and /EY/. We can obtain the the word score of 0.6 by averaging the corresponding GOP scores, which are 0.3 and 0.9 respectively. 

Compared to \cite{lin2021deep}, the proposed mixup is able to generate as many as artificial word samples with a small amount of task-related unlabeled data. Different from \cite{lee2016language,korzekwa2020detection}, where the label of generated training samples is binary(e.g., correct or mispronounced), proposed mixup can generate continuous scores for augmented data.

\subsection{Pronunciation Scorer}
Word pronunciation scorer is designed to mapping acoustic feature to the word score. The Word pronunciation scorer consists of two training steps: (1) pretraining with unlabeled and mixup augmented data and (2) fine-tuning with human labeled data. 
%A amount of GOP scoring data are used to pretrain the scorer, such as task-related unlabeled data and the artificial data by mixup.
%The human labeled data in \cite{zhang2021speechocean762}, used to update the pretrained model.

% During pretraining, two sets of data were used for model training, including task related unlabeled data and augmented data generated by mixup (as described in Section~\ref{ssec:mixup}). Given a word with sequence of frame-level deep features \{$\mathbf{x}_1^\text{d}$, $\mathbf{x}_2^\text{d}$, ...,$\mathbf{x}_T^\text{d}$\}, we first apply nonlinear transformation on each of them, and then add them with corresponding canonical phone embeddings to obtain a sequence of frame-level phonetic features, which is subsequently fed into the 1D-CNN ~\cite{cnn} module. Then a mean-overtime pooling operation is applied over the feature map obtained by 1D-CNN module, to get the word-level representation $\mathbf{h}^\text{d}$. Similarly, $\mathbf{h}^\text{m}$ can be obtained after doing the same operation on the MFCC features. Finally, their concatenation is fed into Eq. (\ref{eq:predict}) to predict word pronunciation score $p$.

During pretraining, two sets of data were used for model training, including task related unlabeled data and augmented data generated by mixup (as described in Section~\ref{ssec:mixup}). Given a word with sequence of frame-level deep features \{$\mathbf{x}_1^\text{d}$, $\mathbf{x}_2^\text{d}$, ...,$\mathbf{x}_T^\text{d}$\}, we first apply nonlinear transformation on each of them, and then add them with corresponding canonical phone embeddings to obtain a sequence of frame-level phonetic features, which is subsequently fed into the 1D-CNN ~\cite{cnn} module. Then a mean-overtime pooling operation is applied over the feature map obtained by 1D-CNN module, to get the word-level representation $\mathbf{h}^\text{d}$. Similarly, $\mathbf{h}^\text{m}$ can be obtained after doing the same operation on the MFCC features. Finally, their concatenation is fed into Eq.({\ref{eq:predict}}) to predict word pronunciation score $p$.
\begin{equation}
\label{eq:predict}
\begin{split}
p =\text{Sigmoid}(\mathbf{W}([\mathbf{h}^\text{d};\mathbf{h}^\text{m}])+\mathbf{b})
\end{split}
\end{equation}  
where the [;] denotes the concatenation of two vectors. The mean square error (MSE) is used as the train objective for model updating, which is defined as:
\begin{equation}
\label{eq:loss}
\begin{split}
\mathcal{L} 
&= \frac{1}{n} \sum_{i=1}^{n} ( y_i^\text{GOP} -  p_i )^2
\end{split}
\end{equation}
where, $n$ denotes the total number of word samples in the pretraining data. $y_i^\text{GOP}$ is the word GOP score of the $i$-th word. 

At fine-tune stage, we adjust the pretrained pronunciation scorer by a small amount of human labeled data. Similar to the process in pretraining, the scorer input consists of MFCC, deep feature and corresponding phoneme sequence. And the word pronunciation score is obtained by Eq. (\ref{eq:predict}).

Different to pretaining, at fine-tune, the $i$-th word score $y_i^\text{Human}$ assigned by human rater are used to supervise the scorer updating. The Eq. (\ref{eq:loss}) can be rewrite as:
\begin{equation}
\label{eq:human}
\begin{split}
\mathcal{L} 
&= \frac{1}{m} \sum_{i=1}^{m} ( y_i^\text{Human} -  p_i )^2
\end{split}
\end{equation}
where $m$ denotes the total number of word samples in the fine-tuning data.

\section{Experimental setup}
\label{sec:exp_setup}
\subsection{Speech Corpus}
\label{ssec:dataset_setup}

% \begin{table}[t!]
% \footnotesize
% \centering
% \caption{Training and testing statistics.}
% \begin{adjustbox}{width=0.3\textwidth,center}
% \begin{tabular}{c|c|c}
%     \toprule 
    
%     Speechocean762  & Train & Test \\
%     \midrule
%     \# of words & 15,849 & 15,967 \\
%     \# of phones & 47,390 & 47,688 \\
    
%     \bottomrule
% \end{tabular}
% \label{table:data}
% \end{adjustbox}
% \end{table}

% \usepackage{multirow}

The ASR training corpus consists of 5,000-hour speech data, including 960-hour native speech from LibriSpeech corpus~\cite{librispeech}, and 4,000-hour non-native private recordings from Bytedance. While the pronunciation scoring data consists of 5,000 English utterances spoken by 250 non-native learners from Speechocean762 \footnote{http://www.openslr.org/101/}. The word pronunciation score in each of them has been labeled by five experts, the median scores were adopted following the score files coming with the database, ranging from 0 to 10. Following the calculation method of previous study~\cite{nancypcc}, the averaged inter-rater agreement is 0.726. In addition, linguistic experts in Bytedance collected a small amount of task-related unlabeled data (e.g., a group of Chinese adults are required to read aloud given English prompts). It is used in proposed mixup for word-level augmented data generation. The statistics of pronunciation scoring data is detailed in Table \ref{table:data}.

%\textcolor{red}{For Speechocean762, multiple inter-rater PCC were calculated between the scores of one rater and the median scores of the rest of all raters is 0.726, which is the upper bound of the scoring performance (expert performance).}

\begin{table}[h!]
\centering
\caption{Data statistics for pronunciation scorer.}
\begin{adjustbox}{width=0.48\textwidth,center}
\begin{tabular}{c|c|c|c}
    \toprule 
    %  &  & & \\
\multirow{1}{*}{} &\multirow{2}{*}{ \textbf{Task-related unlabeled data} }&\multicolumn{2}{c}{ \textbf{Speechocean762} }   \\
 & & \multirow{1}{*}{Train}  & \multirow{1}{*}{Test} \\ 
%    Speechocean762  & Train & Test \\
    \midrule
    \# of words & 50,000 & 15,849 & 15,967 \\
    \# of phones & 168,472  & 47,390 & 47,688 \\
    
    \bottomrule
\end{tabular}
\end{adjustbox}
\label{table:data}
\end{table}

\subsection{Experimental configuration}
\label{ssec:ASR_training}
DFSMN-HMM~\cite{dfsmn} based ASR model was adopted, which consists of 2 convolutional layers, 24 FSMN layers, a bottleneck layer and a feedforward layer. 
39-dimensional MFCC feature was extracted using a 25 ms hamming window with 10 ms shift as the model input. While, the softmax output layer had 5,432 units, representing the senone labels derived from forced-alignment with a GMM-HMM system. 
The deep features were extracted from output of the bottleneck layer with the dimension of 512. While, the phoneme GOP scores were calculated with frame-level senones and their corresponding posteriors. The frame accuracy of the ASR system is 73\%.

%\subsection{Pronunciation Scoring Model Configuration}
%\label{ssec:scoring}

\begin{table}[t!]
  \caption{The architecture of the proposed 1D-CNN-MLP, where $L_{i} = (L_{i-1} + 2*padding -kernel\_size)/stride + 1  $, and $L_0$ is the frame length of one word.}
  \label{table:cnn}
  \small
  \centering
  \begin{adjustbox}{width=0.48\textwidth,center}
  \begin{tabular}{c|c|c|c}
    \toprule
     &\multicolumn{1}{c|}{Layer} & \multicolumn{1}{c|}{Description} &\multicolumn{1}{c}{Output Size} \\
    \midrule
    
    %\begin{tabular}[c]{@{}c@{}}MGB-3 Train + MGB-3 Dev\\ + YouTube %Train\end{tabular}& $\mathcal{A}+\mathcal{B}$ & 61.86 & 81.53 \\ \hline \hline
    
    1 & Input &  \begin{tabular}[c]{@{}c@{}}a sequence of phonetic \\ features  \end{tabular} & 32$\times{L_0}$   \\
    \midrule
         2           & 1D-Convolution      &  \begin{tabular}[c]{@{}c@{}}kernel\_size=3, stride=1,\\ 32 filters   \end{tabular}      & 32$\times{L_1}$         \\
         3           & Batch normalization    &     & 32$\times{L_1}$          \\
         4           & ReLU   &       & 32$\times{L_1}$          \\
         5          & Dropout &  probability: 0.1    & 32$\times{L_1}$        \\
    \midrule
         6           & 1D-Convolution            &  \begin{tabular}[c]{@{}c@{}}kernel\_size=3, stride=1, \\ 32 filters   \end{tabular}   & 32$\times{L_2}$       \\
         7           & Batch normalization    &                 & 32$\times{L_2}$     \\
         8          & ReLU                   &                 & 32$\times{L_2}$         \\
         9           & Dropout                &  probability: 0.1            & 32$\times{L_2}$        \\
        10            &Max pooling             &  kernel\_size=2,stride=2   & 32$\times{L_3}$        \\
    \midrule
        11           & 1D-Convolution            &  \begin{tabular}[c]{@{}c@{}}kernel\_size=1, stride=1, \\ 32 filters   \end{tabular}      & 32$\times{L_4}$       \\
        12            & Batch normalization    &                 & 32$\times{L_4}$     \\
        13          & ReLU                   &                 & 32$\times{L_4}$     \\
        14            & Dropout                &  probability: 0.1            & 32$\times{L_4}$     \\
                   
    \midrule
        15            & Mean pooling           &  average time   & 32$\times$1         \\
        16          & Feature concatenation        & mfcc + deep feature                 & 64$\times$1       \\
    
    \midrule
        17            & Fully connected        & output size: 32                 & 32$\times$1        \\
        18          & Fully connected        &  output size: 1               & 1$\times$1         \\
        19            & Sigmoid                & predicted word score & 1$\times$1         \\
                   
    \bottomrule
 
  \end{tabular}
 \end{adjustbox}
\end{table}

% The pronunciation scoring consists of two steps: pre-training and fine-tuning.
% The model was first pre-trained with the augmented data generated from Speechocean762 training set and the task-related unlabeled data shown in Table \ref{table:data}. The phone-level GOP scores extracted from ASR model were used to form the word-level scores for augmented data.
% Then the model was fine-tuned on Speechocean762 training set with the human-labeled word scores.

% The multi-source model takes sequence of frame-level 39-dim MFCC and 512-dim deep feature as its input. Nonlinear transformations are applied to these features, and output frame-level 32-dim feature vectors, which are added to 32-dim reference phone embeddings to get phonetic features. Subsequently, we use CNN and MLP module to map phonetic features into word score. The detailed configuration is listed in Table \ref{table:cnn}. The Adam optimizing algorithm is chosen to minimize the MSE loss described in Eq. (\ref{eq:loss}), and learning rate is set to 0.002. The word scores of both augmented and human-labeled data are scaled to the interval from 0 to 1 for model training and evaluation.

The multi-source scoring model takes sequence of frame-level 39-dim MFCC and 512-dim deep feature as its input. Nonlinear transformations are applied to these features, and output frame-level 32-dim feature vectors, which are added to 32-dim reference phone embeddings to get phonetic features. Subsequently, we use 1D-CNN and MLP module to map phonetic features into word score. The detailed configuration is listed in Table \ref{table:cnn}. The Adam optimizing algorithm is chosen to minimize the MSE, and learning rate is set to 0.002. The word scores of both augmented and human-labeled data are scaled to the interval from 0 to 1 for model training and evaluation.

\subsection{System Setup}
\label{ssec:baselines}
To valid the proposed ideas, we considered the state-of-the-art method as our baseline.

\leftmargini=5mm
\begin{itemize}
\item{\textbf{No-pretrain}}: the scoring model is trained on the human-labeled data directly. No pretraining process is applied;
\item{\textbf{Real-pretrain}}: we calculate GOP score of each word in task-related unlabeled data, and use it to pretrain word-level scorer, which is then fine-tuned with the labeled data in training set shown in Table~\ref{table:data}. This pretrain and fine-tune strategy achieved the highest PCC for sentence scoring in the recent study~\cite{lin2021deep}, hence we apply it for word-level scoring and treat it as a strong baseline;

% the word-level scoring system with whole training set, including both labeled and unlabeled data, for pretraining~\cite{lin2021deep}. The model is then fine-tuned with the labeled data in training set;
\item{\textbf{Mixup-pretrain}}: The pretraining is first preformed on the task-related unlabeled data and the artificial data generated by mixup. The model is then fine-tuned with the labeled data in training set. 
\end{itemize}

For fair comparison, both baselines and proposed system shared the same ASR and scoring model structures.

\section{Results and analyses}
\label{sec:res}

To assess the system performance, the Pearson correlation coefficient (PCC) between machine predicted scores and human scores was used in our experiments. A higher PCC indicates a better system performance.

\subsection{Validate the Configurations of Proposed Method}
\label{ssec:self_config}

In this section, we examined two configurations of proposed method, including the effects of mixup data size and multi-source information for predicting word-level pronunciation quality.

\begin{figure}[tb]
\centering
\includegraphics[width=.5\textwidth]{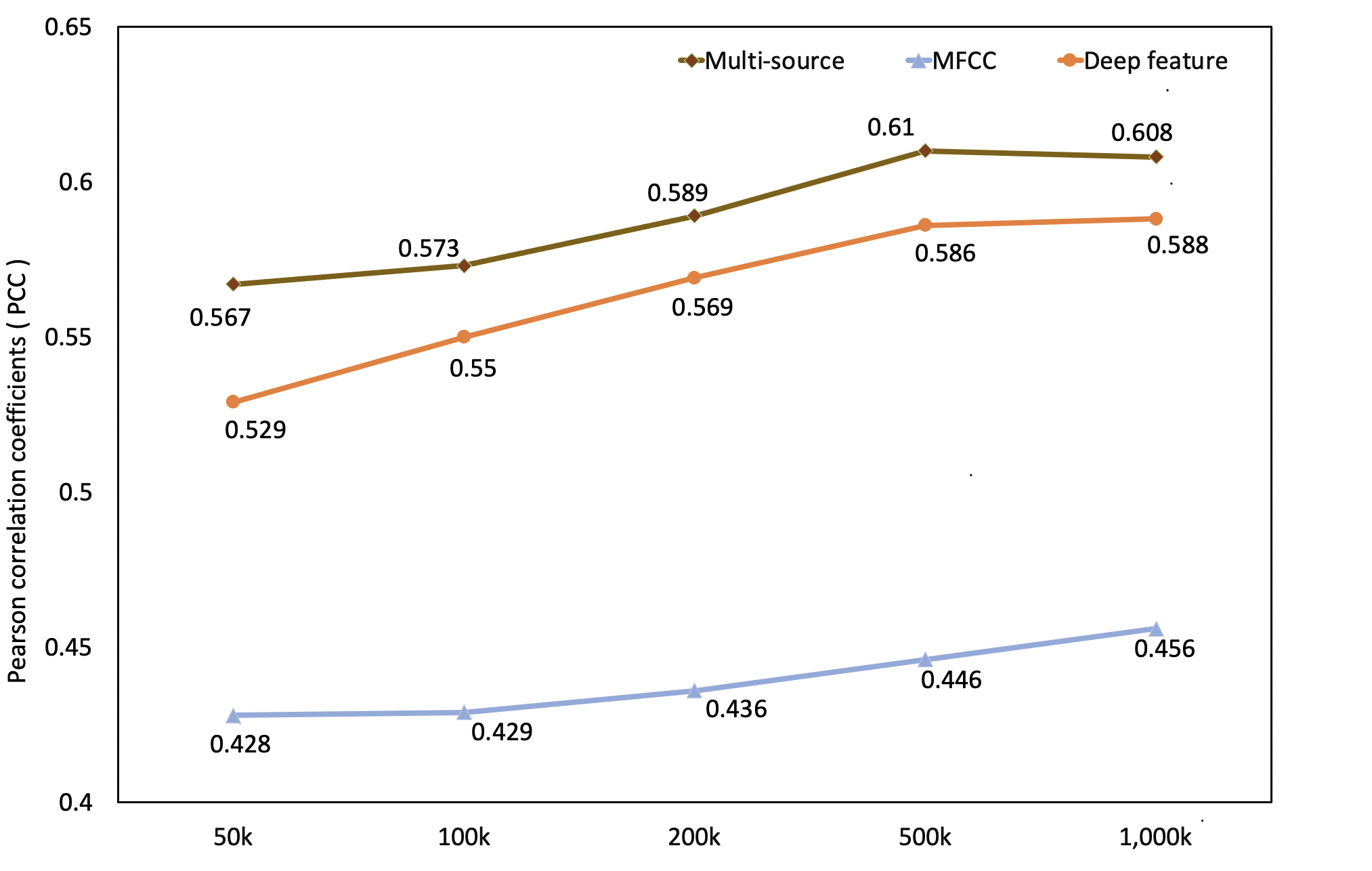}
\caption{Performance of different amount of augmented data. The augmented data consists of both real unlabeled data (50k words) and artificial data generated with mixup. Take augmented data size of 500k as an example, it is formed with 50k real unlabeled data and 450k mixup data. Augmented data size of 50k indicates no mixup data.}
\label{fig:mixup_data}
\end{figure}

\begin{table}[t!]
\vspace{1pt}
  \centering
  \caption{Comparison of different methods. Mixup-pretrain is pretrained with 50k real unlabeled data and 450k artificial augmented word created by mixup.}
\begin{tabular}{lcc}
    \toprule
    \textbf{Methods}  & \textbf{Real unlabeled data}  & \textbf{PCC} \\
    \midrule
  % \emph{Human}  &  - & 0.726 \\
  %  GOP    &  - & 0.44 \\ 
    No-pretrain &  - & 0.507 \\
    \midrule
    \multirow{4}{*}{Real-pretrain} &  50k & 0.567 \\
      &  100k &  0.570 \\
      &  200k & 0.586 \\
      &  500k & 0.612 \\
    \midrule
    \textbf{Mixup-pretrain}  &  50k & \textbf{0.610} \\
    %\midrule
    % \midrule
    % \emph{Human}  &  - & 0.726 \\
    
    \bottomrule
\end{tabular}
\vspace{0pt}
%%%%这个地方描述是否引起人的混淆？
%\caption{Comparison of different methods. Real unlabeled data is the task-related non-native data without human label, which will be used for data augmentation in real- and mixup-pretrain methods. Mixup-pretrain is pretrained with 50k real unlabeled data and 450k artificial augmented word created by mixup.}
\label{table:compare_baseline}
\end{table}

\subsubsection{Effect of Mixup Data Size}
\label{ssec:data_size}

We first compare the performance of the word-level scoring over different size of augmented data, which consists of real unlabeled data (50k words) and artificial data generated with mixup, e.g. 50k, 100k, 200k, 500k and 1,000k words. Figure~\ref{fig:mixup_data} presents the PCC results with different amount of augmented data. It is observed that the system performance improves as the augmented data size increases for all three features. When the augmented data size reaches 500k, the best performance is achieved, with the PCC of 0.610. After that, we observe the PCC plateaus out for multi-source and deep features. Therefore, we set the augmented data size as 500k words (50k unlabeled data + 450k mixup data) in the rest of the experiments.

% \begin{table}[t!]
% \vspace{1pt}
%   \centering
% \begin{tabular}{lcc}
%     \toprule
%     \textbf{Methods}  & \textbf{Real unlabeled data}  & \textbf{PCC} \\
%     \midrule
%   % \emph{Human}  &  - & 0.726 \\
%   %  GOP    &  - & 0.44 \\ 
%     No-pretrain &  - & 0.507 \\
%     \midrule
%     \multirow{4}{*}{Real-pretrain} &  50k & 0.567 \\
%       &  100k &  0.570 \\
%       &  200k & 0.586 \\
%       &  500k & 0.612 \\
%     \midrule
%     \textbf{Mixup-pretrain}  &  50k & \textbf{0.610} \\
%     %\midrule
%     % \midrule
%     % \emph{Human}  &  - & 0.726 \\
    
%     \bottomrule
% \end{tabular}
% \vspace{0pt}
% %\caption{Comparison of different methods. Real unlabeled data is the task-related non-native data without human label, which will be used for data augmentation in real- and mixup-pretrain methods. Mixup-pretrain is pretrained with 50k real unlabeled data and 450k artificial augmented word created by mixup.}
% \caption{Comparison of different methods. Mixup-pretrain is pretrained with 50k real unlabeled data and 450k artificial augmented word created by mixup.}

% \label{table:compare_baseline}
% \end{table}

\subsubsection{Effect of Multi-Source Information for Scoring}
\label{ssec:multi-source}

Then, we examined word-level pronunciation scoring performance over different input, e.g. MFCC, deep and multi-source feature. As shown in Figure~\ref{fig:mixup_data}, it is observed that deep feature consistently outperforms the MFCC for pronunciation scoring in all the tests. This observation confirms that deep feature is more discriminative than MFCC due to utilizing extra large amount of non-native data for its feature extractor training. The performance is further improved by multi-source feature, which utilizing both MFCC and deep features as system input. The results imply that MFCC and deep features are complementary to each other, and their combination in hidden space is helpful for pronunciation scoring.

\subsection{Comparison with Baselines}
\label{ssec:comparison_baseline}

In this section, we validate the proposed mixup by comparing against baselines.
To focus on comparison of different pretraining methods, all the systems take multi-source feature as their input. %Because Figure 2 shows that this feature setup achieves the highest PCC.   
Table~\ref{table:compare_baseline} presents the PCC results from different methods. Benefit from the real unlabeled non-native data, all the real- and mixup-pretrain methods outperform the no-pretrain counterpart. This confirms the effectiveness of the data augmentation and pretrainining method for pronunciation scoring.

Then we compare the proposed mixup-pretrain against real-pretrain baseline. Since the task is defined in a low-resource scenario, we fix the size of real unlabelled data as 50k for the proposed method. Experimental results indicate that the proposed mixup-pretrain method performs better than the real-pretrain baseline trained with the same amount of unlabelled data when using 50k real unlabelled data. To achieve comparable performance of real-pretrain with the proposed mixup-pretrain, we increase the amount of real unlabeled data from 50k to 500k. From Table~\ref{table:compare_baseline}, we clearly see that the system performance improves as the unlabeled data size increases. The results show that, the real-pretrain baseline achieves similar performance to the proposed method when using 500k real unlabelled data, which is 10 times
of proposed method. It suggests that the proposed mixup
method offers a clear advantage over baseline under limited data condition.

% Then we compare the proposed mixup-pretrain against real-pretrain baseline. By using the same amount of real unlabeled data (50k words), the mixup-pretrain achieves the PCC of 0.610, which outperforms the baseline trained with the same amount of unlabeled data (PCC of 0.567). We then examine the performance of real-pretrain by increasing the size of unlabeled data, e.g. 100k, 200k and 500k words respectively. From Table~\ref{table:compare_baseline}, we clearly see that the system performance improves as the unlabeled data size increases, and achieves the best PCC of 0.612.
% It is observed that the mixup-pretrain achieves the comparable performance of real-pretrain using 500k real unlabeled data, which is 10 times of proposed method.
% The results suggest that the proposed mixup method offers a clear advantage over baseline under limited non-native data condition. 

\section{Conclusion}
In this paper, phone-level mixup is proposed to make full use of smaller unit (e.g., phone) to creates arbitrary bigger unit (e.g., word) with various pronunciation scores. Experiments conducted on Speechocean762 show that proposed augmented data is complementary to the real non-native data for training word-level pronunciation scorer. Moreover, feature combination under the proposed training framework brings extra improvement. In the future, we will extend current phone-level mixup to word-level mixup for improving sentence-level scoring. Meanwhile, we will also investigate whether similar improvement could be achieved, when the proposed data augmentation method is applied to attention-based pronunciation scorers \cite{korzekwa2021weakly,lin2020automatic}.

\clearpage
\newpage

\bibliographystyle{IEEEtran}
\bibliography{./mybib}

% Generated by IEEEtran.bst, version: 1.14 (2015/08/26)
\begin{thebibliography}{10}
\providecommand{\url}[1]{#1}
\csname url@samestyle\endcsname
\providecommand{\newblock}{\relax}
\providecommand{\bibinfo}[2]{#2}
\providecommand{\BIBentrySTDinterwordspacing}{\spaceskip=0pt\relax}
\providecommand{\BIBentryALTinterwordstretchfactor}{4}
\providecommand{\BIBentryALTinterwordspacing}{\spaceskip=\fontdimen2\font plus
\BIBentryALTinterwordstretchfactor\fontdimen3\font minus
  \fontdimen4\font\relax}
\providecommand{\BIBforeignlanguage}[2]{{%
\expandafter\ifx\csname l@#1\endcsname\relax
\typeout{** WARNING: IEEEtran.bst: No hyphenation pattern has been}%
\typeout{** loaded for the language `#1'. Using the pattern for}%
\typeout{** the default language instead.}%
\else
\language=\csname l@#1\endcsname
\fi
#2}}
\providecommand{\BIBdecl}{\relax}
\BIBdecl

\bibitem{ellis1994study}
R.~Ellis and R.~R. Ellis, \emph{{The study of second language
  acquisition}}.\hskip 1em plus 0.5em minus 0.4em\relax Oxford University,
  1994.

\bibitem{meng2009developing}
H.~Meng, ``{Developing speech recognition and synthesis technologies to support
  computer-aided pronunciation training for Chinese learners of English},'' in
  \emph{Proceedings of the 23rd Pacific Asia Conference on Language,
  Information and Computation, Volume 1}, 2009, pp. 40--42.

\bibitem{franco1997automatic}
H.~Franco, L.~Neumeyer, Y.~Kim, and O.~Ronen, ``{Automatic pronunciation
  scoring for language instruction},'' in \emph{ICASSP}.\hskip 1em plus 0.5em
  minus 0.4em\relax IEEE, 1997.

\bibitem{witt2000phone}
S.~M. Witt and S.~J. Young, ``{Phone-level pronunciation scoring and assessment
  for interactive language learning},'' \emph{Speech communication}, vol.~30,
  no. 2-3, pp. 95--108, 2000.

\bibitem{zheng2007generalized}
J.~Zheng, C.~Huang, M.~Chu, F.~K. Soong, and W.-p. Ye, ``Generalized segment
  posterior probability for automatic mandarin pronunciation evaluation,'' in
  \emph{2007 IEEE international conference on acoustics, speech and signal
  processing (ICASSP)}, vol.~4.\hskip 1em plus 0.5em minus 0.4em\relax IEEE,
  2007, pp. IV--201.

\bibitem{hu2015improved}
W.~Hu, Y.~Qian, F.~K. Soong, and Y.~Wang, ``{Improved mispronunciation
  detection with deep neural network trained acoustic models and transfer
  learning based logistic regression classifiers},'' \emph{Speech
  Communication}, vol.~67, pp. 154--166, 2015.

\bibitem{shi2020context}
J.~Shi, N.~Huo, and Q.~Jin, ``Context-aware goodness of pronunciation for
  computer-assisted pronunciation training,'' in \emph{Proc. Interspeech},
  2020.

\bibitem{li2017improving}
W.~Li, N.~F. Chen, S.~M. Siniscalchi, and C.-H. Lee, ``{Improving
  Mispronunciation Detection for Non-Native Learners with Multisource
  Information and LSTM-Based Deep Models.}'' in \emph{Proc. Interspeech}, 2017,
  pp. 2759--2763.

\bibitem{leung2019cnn}
W.-K. Leung, X.~Liu, and H.~Meng, ``{CNN-RNN-CTC based end-to-end
  mispronunciation detection and diagnosis},'' in \emph{ICASSP 2019-2019 IEEE
  International Conference on Acoustics, Speech and Signal Processing
  (ICASSP)}.\hskip 1em plus 0.5em minus 0.4em\relax IEEE, 2019, pp. 8132--8136.

\bibitem{yan2020end}
B.-C. Yan, M.-C. Wu, H.-T. Hung, and B.~Chen, ``{An end-to-end mispronunciation
  detection system for L2 English speech leveraging novel anti-phone
  modeling},'' in \emph{Proc. Interspeech}, 2020, pp. 3032--3036.

\bibitem{feng2020sed}
Y.~Feng, G.~Fu, Q.~Chen, and K.~Chen, ``{SED-MDD: Towards sentence dependent
  end-to-end mispronunciation detection and diagnosis},'' in \emph{ICASSP
  2020-2020 IEEE International Conference on Acoustics, Speech and Signal
  Processing (ICASSP)}.\hskip 1em plus 0.5em minus 0.4em\relax IEEE, 2020, pp.
  3492--3496.

\bibitem{wei2009new}
S.~Wei, G.~Hu, Y.~Hu, and R.-H. Wang, ``{A new method for mispronunciation
  detection using support vector machine based on pronunciation space
  models},'' \emph{Speech Communication}, vol.~51, no.~10, pp. 896--905, 2009.

\bibitem{qian2012use}
X.~Qian, H.~Meng, and F.~K. Soong, ``{The use of DBN-HMMs for mispronunciation
  detection and diagnosis in L2 English to support computer-aided pronunciation
  training},'' in \emph{Thirteenth Annual Conference of the International
  Speech Communication Association}, 2012.

\bibitem{kyriakopoulos2020automatic}
K.~Kyriakopoulos, K.~Knill, and M.~Gales, ``{Automatic detection of accent and
  lexical pronunciation errors in spontaneous non-native English speech},'' in
  \emph{Proc. Interspeech}, 2020.

\bibitem{lee2013pronunciation}
A.~Lee and J.~Glass, ``{Pronunciation assessment via a comparison-based
  system},'' in \emph{Speech and Language Technology in Education}, 2013.

\bibitem{lee2016language}
A.~Lee \emph{et~al.}, ``{Language-independent methods for computer-assisted
  pronunciation training},'' Ph.D. dissertation, Massachusetts Institute of
  Technology, 2016.

\bibitem{korzekwa2021weakly}
D.~Korzekwa, J.~Lorenzo-Trueba, T.~Drugman, S.~Calamaro, and B.~Kostek,
  ``{Weakly-supervised word-level pronunciation error detection in non-native
  English speech},'' pp. 4408--4412, 2021.

\bibitem{cincarek2009automatic}
T.~Cincarek, R.~Gruhn, C.~Hacker, E.~N{\"o}th, and S.~Nakamura, ``{Automatic
  pronunciation scoring of words and sentences independent from the
  non-native’s first language},'' \emph{Computer Speech \& Language},
  vol.~23, no.~1, pp. 65--88, 2009.

\bibitem{yu2015using}
Z.~Yu, V.~Ramanarayanan, D.~Suendermann-Oeft, X.~Wang, K.~Zechner, L.~Chen,
  J.~Tao, A.~Ivanou, and Y.~Qian, ``{Using bidirectional LSTM recurrent neural
  networks to learn high-level abstractions of sequential features for
  automated scoring of non-native spontaneous speech},'' in \emph{ASRU}.\hskip
  1em plus 0.5em minus 0.4em\relax IEEE, 2015, pp. 338--345.

\bibitem{chen2018end}
L.~Chen, J.~Tao, S.~Ghaffarzadegan, and Y.~Qian, ``{End-to-end neural network
  based automated speech scoring},'' in \emph{2018 IEEE International
  Conference on Acoustics, Speech and Signal Processing (ICASSP)}.\hskip 1em
  plus 0.5em minus 0.4em\relax IEEE, 2018, pp. 6234--6238.

\bibitem{lin2020automatic}
B.~Lin, L.~Wang, X.~Feng, and J.~Zhang, ``{Automatic Scoring at
  Multi-Granularity for L2 Pronunciation.}'' in \emph{Proc. Interspeech}, 2020,
  pp. 3022--3026.

\bibitem{lin2021deep}
B.~Lin and L.~Wang, ``{Deep Feature Transfer Learning for Automatic
  Pronunciation Assessment},'' \emph{Proc. Interspeech}, pp. 4438--4442, 2021.

\bibitem{zhang2021speechocean762}
J.~Zhang, Z.~Zhang, Y.~Wang, Z.~Yan, Q.~Song, Y.~Huang, K.~Li, D.~Povey, and
  Y.~Wang, ``speechocean762: An open-source non-native english speech corpus
  for pronunciation assessment,'' in \emph{Proc. Interspeech}, 2021.

\bibitem{zhao2018l2}
G.~Zhao, S.~Sonsaat, A.~Silpachai, I.~Lucic, E.~Chukharev-Hudilainen, J.~Levis,
  and R.~Gutierrez-Osuna, ``L2-arctic: A non-native english speech corpus.'' in
  \emph{Proc. Interspeech}, 2018, pp. 2783--2787.

\bibitem{nancy}
N.~F. Chen, D.~Wee, R.~Tong, B.~Ma, and H.~Li, ``{Large-scale characterization
  of non-native Mandarin Chinese spoken by speakers of European origin:
  Analysis on iCALL},'' \emph{Speech Communication}, vol.~84, pp. 46--56, 2016.

\bibitem{korzekwa2020detection}
D.~Korzekwa, R.~Barra-Chicote, S.~Zaporowski, G.~Beringer, J.~Lorenzo-Trueba,
  A.~Serafinowicz, J.~Droppo, T.~Drugman, and B.~Kostek, ``{Detection of
  lexical stress errors in non-native (l2) english with data augmentation and
  attention},'' in \emph{Proc. Interspeech}, 2021.

\bibitem{mixup}
H.~Zhang, M.~Ciss{\'{e}}, Y.~N. Dauphin, and D.~Lopez{-}Paz, ``mixup: Beyond
  empirical risk minimization,'' in \emph{6th International Conference on
  Learning Representations, {ICLR}}, 2018.

\bibitem{cnn}
Y.~Kim, ``{Convolutional Neural Networks for Sentence Classification},''
  \emph{Proceedings of the 2014 Conference on Empirical Methods in Natural
  Language Processing}, 08 2014.

\bibitem{librispeech}
V.~Panayotov, G.~Chen, D.~Povey, and S.~Khudanpur, ``{Librispeech: an asr
  corpus based on public domain audio books},'' in \emph{2015 IEEE
  international conference on acoustics, speech and signal processing
  (ICASSP)}.\hskip 1em plus 0.5em minus 0.4em\relax IEEE, 2015, pp. 5206--5210.

\bibitem{nancypcc}
H.~Zhang, K.~Shi, and N.~Chen, ``Multilingual speech evaluation: Case studies
  on english, malay and tamil,'' \emph{Proc. Interspeech}, pp. 4443--4447,
  2021.

\bibitem{dfsmn}
S.~Zhang, M.~Lei, Z.~Yan, and L.~Dai, ``{Deep-FSMN for large vocabulary
  continuous speech recognition},'' in \emph{ICASSP 2018-2018 IEEE
  International Conference on Acoustics, Speech and Signal Processing
  (ICASSP)}.\hskip 1em plus 0.5em minus 0.4em\relax IEEE, 2018, pp. 5869--5873.

\end{thebibliography}

\end{document}